# Embedded real-time monitoring using SystemC in IMA network

N. Ahamada[1], Z. Aloui[1], J. Denoulet[1], F. Pierre[2], M. Rayrole[2], M. Gatti[2], B. Granado[2]

1 – Sorbonne Universités, UPMC Univ Paris 06, UMR7606, LIP6, F70005 Paris, France, 2 – Thales Avionics, 19 Avenue Morane Saulnier, 78140 Velizy-Villacoublay

## Abstract

Avionics is one kind of domain where prevention prevails. Nonetheless fails occur. Sometimes due to pilot misreacting, flooded in information. Sometimes information itself would be better verified than trusted. To avoid some kind of failure, it has been thought to add, in midst of the ARINC664 aircraft data network, a new kind of monitoring.

## 1 Introduction

It is well known that avionics is a very restricting domain for obvious safety reasons. Along with miniaturization comes the idea of integration. More functionality on one spot requires a good management of privacy and congestion on shared platforms. This is why determinism is one of the keywords of avionics works. This led to protocols like ARINC653[1] assuring that, multitask embedded programs respect a predictable policy applied by the operating system (OS). Another key protocol is ARINC664, which guarantees that multiple communicating systems efficiently share the network. These two protocols are pillars of the Integrated Modular Architecture (IMA) concept [2]. IMA concept consists of multitask module hosting ARINC653 OS, interconnected with ARINC664 data network. Compared to federated avionics architecture, it considerably reduces the overall weight and power consumption for aircraft, reduces development expenses and design cycle times as well as maintenance costs. With the intention to step forward with this concept, the CORAC (The Council for Civil Aeronautics Research) develops a technological demonstration platform (PDT) called Extended Modular Avionic (AME) [3]. Therefore, as partner of the project, we work on a project dedicated to monitor the system.

In this paper, we propose to use a new kind of monitoring based on embedded simulation. This simulator implemented in SystemC, monitors data traffic generated by key aviations in order to detect suspicious behavior such as missing data, unexpected communication of simply incoherent data. In the next sections, we first introduce our method and its related tools, most notably the SystemC language and the modifications required to make this language compliant with avionic constraints. Then we introduce a use case which will validate our method. Finally, we will conclude..

## 2 Method

### 2.1 Method presentation

Considering the predictability and determinism of applications software ruled by the protocol ARINC653 and their windows of communication in ARINC664, one can predict part of the aircraft data traffic. Some verification within the communication protocol already exist concerning the integrity of the data transport but none can analyze the content itself to determine whether one or another application is really supposed to send a value, or if a communication disappeared or if a value is simply incoherent. Obviously simulating the whole communication flow to determine if it is coherent would be too much time expensive in simulation. The idea is to target specific applications, or specific suspect behaviors (missing material, erroneous values) we could watch over during the flight. Knowing what we're looking for, we could then create a simplified functional timed model of applications as communication providers. On the basis of ARINC664 and ARINC653 configurations values (major frame, bandwidth allocation gap ..), we could predict communication by simulation and compare it with the real traffic. A privileged place to implement a model simulation is the switch modules where the CPU only manages message traffic and have available time.

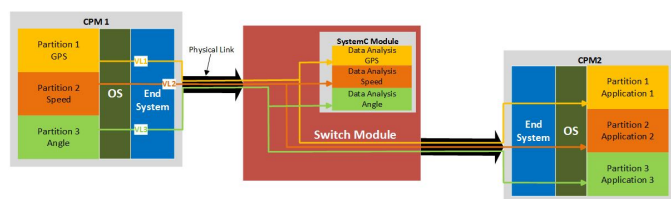

**Figure 1: Monitoring methodology**

In Figure 1, we present the principle of our method. We consider an avionic architecture featuring core processing modules (CPM) implementing several applications and generating data traffic and avionics switch modules (ASM) which route data packets to their destination CPM.

As an example, CPM1 in Figure 1 features three partitions, each one hosting an application dedicated respectively to GPS, Speed, and Angle estimation. Through an ARINC664 communication End System, data generated by these applications are sent through several Virtual Links (VL) of the data network. While performing data traffic management, the ASM also implements a simulator that runs a timed model of the expected communication traffic, considering the OS and network parameters. The simulator performs two types of verification: temporal consistency which checks whether communication occurs at the expected time, according to the system scheduling, and data consistency which analyses N consecutive data values to determine if their evolution is coherent or if an error can be assumed to have occurred.

To achieve such a goal, we have chosen the SystemC [4] language as an appropriate candidate to model as well software (application) and hardware system (processors and communication modules) under time constraints (defined by ARINC653 and ARINC664). The next subsections present the SystemC language specifications, as well as SystemCASS, a SystemC simulation kernel that can meet avionics requirements. We finally present the modifications we introduced to the SystemCASS kernel to meet these constraints.

## 2.2 SystemC

SystemC is a C++ class library based on object-oriented design concept (OOD) providing common Hardware Description Language (HDL) features. As such, it allows hardware description alongside with software development. The concurrency of hardware behavior is simulated by the way simulation time is being managed by the simulator.

Hardware components are modeled using the sc_module class and are interconnected to each other with sc_port class objects. Module internal registers are represented by sc_signals, and module behavior by processes, which can be described as functions triggered by the update of ports or signals that are registered in a sensitivity list. A SystemC program usually consists in an elaboration phase where all the elements of the described system are declared and assembled, and where all processes are listed. Then comes the simulation phase, which is initiated by the sc_start method, which is a function of the simulator. Finally, the cleanup phase ends simulation, by cleaning objects and structures.

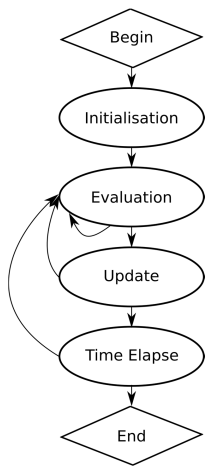

Figure 2: SystemC flow

The role of a SystemC simulator is to manipulate the timestamp to simulate the concurrency of hardware behavior. It determines in which order processes must be executed, and when values of ports and signals must be updated. The Accelera Systems Initiative (ASI) provides an event driven simulator with the language library. The simulator operates according to Figure 2.

During initialization, all processes are put in a state of being executed, which is done at the beginning of the simulation phase. It features three steps: *Evaluation*, in which the simulator checks which processes must be executed, according of their sensitivity list. The simulator then executes these processes. When this is done, the second step, *Update*, updates the values of ports/signals according to the previous processes executions. If signal or ports updates trigger a process sensitivity list again, then we go back to the evaluation step. When no process is triggered anymore, the simulation timestamp is updated in the *Time Elapse* step.

The ASI simulator, as it is implemented, features memory dynamicity, which avionic constraints don't allow. Furthermore,

process scheduling at each timestamp is dynamic and non-deterministic [5]. This doesn't affect the result of the simulation, but can be an issue in an avionic context, considering execution time.

## 2.3 SystemCASS

SystemCASS (SystemC Accurate System Simulator) [6] is a SystemC simulator that establishes a static scheduling of processes, which is made at the start of simulation, based on the considered design (Fig.3).

To do so, SystemCASS requires describing all component models as CFSM (Communicating Finite State Machine) using a CABA (Cycle Accurate Bit Accurate) abstraction level. Furthermore, a single clock must drive all modules.

SystemCASS modules can include three types of processes:

*Transition:* triggered by the clock rising edge, it sets the new values of registers, depending on their actual values as well as input port values.

*Moore Generation:* triggered by the clock falling edge, it sets the new values of output ports, depending on register values only.

*Mealy Generation:* triggered by the clock falling edge, it determines the new values of output ports depending on register values and input port values.

When calling the sc_start method, SystemCASS creates depending graphs that generate the static scheduling of processes, which will be used throughout the simulation phase.

This implementation ensures a deterministic behavior of the simulation. Which is why SystemCASS is more suitable to avionic constraints than a dynamic event driven simulator. As we use gcc compiler, SystemCASS current implementation features dynamic memory allocation during the creation of the depending graph after the elaboration phase, and right before the simulation phase. So we have worked to remove these dynamic allocation. To do so, we have first used a static version of gcc compiler and second we have identified in run-time all the encountered dynamic memory allocations and replaced it with static memory allocations.

# 3 Case study

## 3.1 Case study presentation

To realize a proof-of-concept scenario, we use PolyORB Kernel (POK), a partitioned operating system compliant with ARINC653 avionic standard [7]. POK ensures enforcement of safety and security requirements at run-time. Along with the operating system, POK also provides some example of avionics applications. One of these applications is the Flight Management (see Figure 3) that handles speed, angle and GPS coordinate.

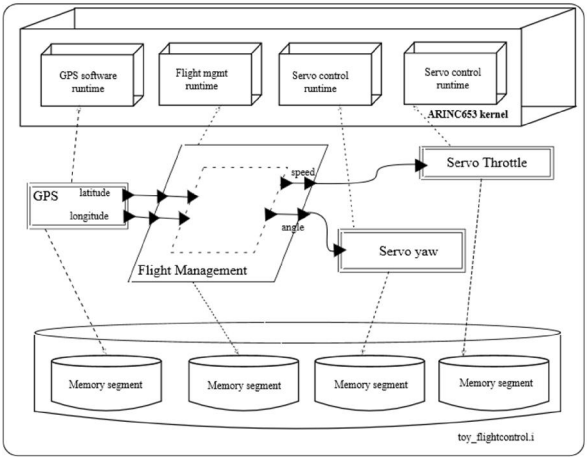

Figure 3: POK Flight Management Application

Our testbed system will be composed of two parts, transmission and reception. This testbed is represented in Figure 4.

The transmission part has two subsystems: the first one, **POK OS Application** generates application data (GPS, speed, angle) while the second one, **End System ARINC664**, encapsulates data to be compliant with the ARINC 664 standard.

On the other hand, the reception part, **SystemC Module Frame Analyzer**, verifies the received ARINC664 frames by performing data consistency based on physical variation laws and temporal consistency by simulating the behavior of the transmission part.

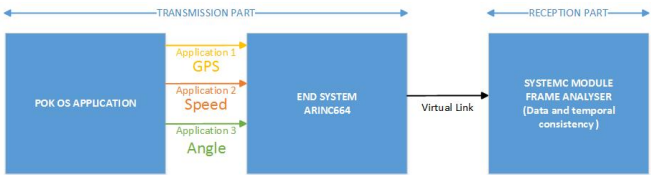

Figure 4: Testbed Structure

## 3.2   Transmission part

ARINC653 guarantees space partitioning (meaning that memory of partition is protected) and also guarantees time partitioning (meaning that only one partition at a time is executed). These properties are also ensured in POK OS. As can be seen in Figure 5, POK will manage three partitions (one for each application) and generates the applications data (speed, angle and GPS).

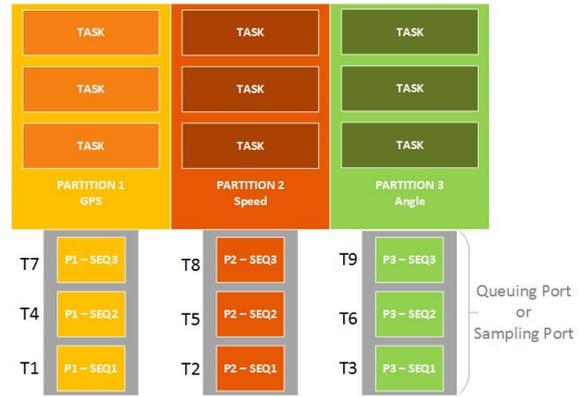

Figure 5: Data Generation and Space Partitionning

The execution of each partition is handled by a static scheduler (as we can see in Figure 6) and is defined by the system integrator. Each partition (P1, P2 and P3) has a set of execution windows (T1, T2, T3) and this set of windows is repeated in time (T4, T5, T6 and so on…) and at the same order, which guarantees that each partition has access to the system resources once in a MAF (Major Frame).

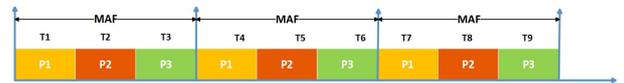

Figure 6: Partitionning Scheduling

Once that data is generated by POK, they are put in the Queuing Port or Sampling Port and are then sent to the End System with the order defined by the scheduler. Queuing Port can be seen as a buffer and the Sampling Port as a FIFO. The End System then encapsulates the data in an ARINC664 frame with the specification of the Virtual Link (BAG, Frame Size, Jitters) that has been defined by the system integrator (see Figure 7). A Virtual Link defines an unidirectional logical connection from one source End-system to one or several destination End-System(s). Each partition has a dedicated Virtual Link (VLi is dedicated to the data of the Partition i).

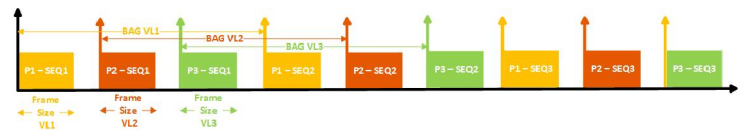

Figure 7: ARINC 664 Frame at the Output of the End System

## 3.3   Reception part

The reception part is a SystemC module that analyzes the ARINC664 frames coming from the transmitter part. It performs data and temporal consistency.

The data consistency consists in analyzing the payload of the ARINC664 frame that contains data of each application (GPS, Speed, angle). In order to do so, a verification of the physical variation law between two data values T and T+1 for each application is performed. For example, the verification of the value of the partition P1-SEQ1 and P1-SEQ2 is performed as shown in Figure 8.

Page 3 of 4

7/20/2015

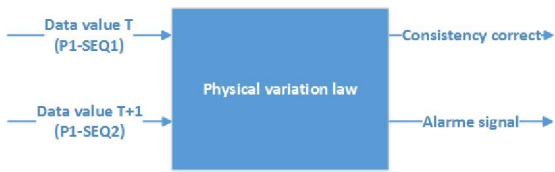

Figure 8: Data Consistency

On the other hand, temporal consistency consists in verifying that the execution order of each partition is consistent with the scheduling defined by the transmitter part Figure 9 shows an example of the temporal consistency verification.

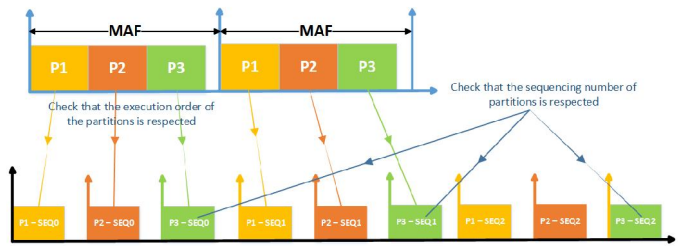

Figure 9: Temporal Consistency Verification

## 4  Conclusion

In this article we have presented a new method to monitor in real time temporal consistency of data circulating through ARINC664 frames. The goal of this monitoring is to checks whether communication occurs at the expected time, according to the system scheduling, and to validate data consistency. To realize this monitoring we use SystemC language and SystemCASS simulator to simulate a timed model of a part of the avionic system.

To validate our system, we are on going to construct a demonstrator based on two Qoriq T2080 design board which has a PowerPC E6500 processor. One board will perform the transmitter part (CPM module) and the second board will perform the reception part (ASM module) as resume the Figure 10.

Each Qoriq T2080 Board (Transmission and Reception) will host the operating system POK and will integrate the SystemC engine. In transmission part (CPM) POK OS will handle the flight management application (GPS, Speed and Angle) and it the same time handle the End System ARINC 664 module that will be developed in SystemC. In Reception part (ASM) POK will handle the SystemC module that performs data and temporal consistency.

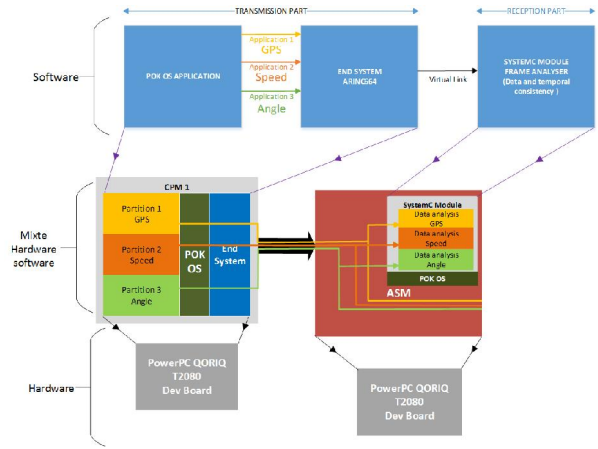

Figure 10: T2080 demonstrator

## Contact Information

Bertrand Granado
Laboratoire LIP6 UMR7606
Université Pierre et Marie Curie
BC 167, Tour 24/25 - 5ieme Etage
4 place jussieu
75252 Paris Cedex 05

Tél : 33 (0)1 44 27 96 33
Email: bertrand.granado@lip6.fr
Website: http://www.lip6.fr